\documentclass[letterpaper, 10pt, conference]{ieeeconf}

\IEEEoverridecommandlockouts                              

\usepackage{epsfig}
\usepackage{amsmath}
\usepackage{amssymb}
\usepackage[T1]{fontenc}
\usepackage{ifthen}
\usepackage{times}
\fontfamily{ptm}\selectfont

\begin{document}

\title{\LARGE \bf Serially Concatenated IRA Codes}

\author{Taikun Cheng, Krishnamoorthy Sivakumar, and Benjamin J. Belzer%
\thanks{The authors are with the School of Electrical Engineering and Computer Science,
        Washington State University, P.O. Box 642752, Pullman, WA 99164-2752, USA
        {\tt\small tcheng,belzer,siva@eecs.wsu.edu}}%
}

\maketitle
\thispagestyle{empty}
\pagestyle{empty}

\begin{abstract}
We address the error floor problem of low-density parity
check (LDPC) codes on the binary-input additive white Gaussian
noise (AWGN) channel,
by constructing
a serially concatenated code consisting of two systematic
irregular repeat-accumulate (IRA) component codes
connected by an interleaver.  The interleaver is designed
to prevent stopping-set error events in one of the IRA codes from
propagating into stopping set events of the other code.  
Simulations with two 128-bit rate 0.707 IRA
component codes show that the proposed
architecture achieves a much lower error floor at higher
SNRs, compared to a 16384-bit rate 1/2 IRA code, but 
incurs an SNR penalty of
about 2 dB at low to medium SNRs.  Experiments indicate that
the SNR penalty can be reduced at larger blocklengths. 
\end{abstract}

\section{Introduction}

LDPC codes, introduced by Gallager in the early 1960s \cite{Gallager},
have received great interest since researchers in the late 1990s and
early 2000s (\cite{MacKay,RU1,RU2}) showed that they can perform within
less than 0.1dB of the Shannon limit for a number of important communication
channels, including the binary erasure channel and the binary-input AWGN
channel.  However, for the above-cited codes, near-capacity performance 
typically holds only above bit error rates (BERs) of $10^{-5}$ or  $10^{-6}$; 
at lower BERs, the nearly vertical (and highly negative) slope of the 
BER vs. SNR curve levels off into an ``error floor'' with a smaller magnitude slope.

As there are several important applications that require BERs of $10^{-12}$ or lower
(e.g., mass storage, broadband satellite communications),
a number of recent publications have proposed LDPCs specially designed to reduce
the error floor.  IRA codes, introduced in \cite{McEliece} by Jin, Khandekar, and McEliece,
feature a section $H_2$ of the parity check matrix $H$ that contains only 
weight-two columns (except for one weight-1 column), and consists of 
``1''s down the main diagonal and the diagonal just below it.  A lemma proved in
\cite{Ryan} shows that if the $H_2$ section contains all the weight two columns of
$H$, then it helps lower the error floor because $H_2$ contains
the maximum number of degree-two variable nodes without a cycle among them.
Extended IRA (e-IRA) codes, introduced in \cite{Ryan}, are a generalization of
systematic IRA codes wherein the remaining section (``$H_1$'') of the $H$ matrix assumes a more 
general form; design rules for lowering the error floor of e-IRA codes by optimizing
the degree distributions of $H_1$ are given in \cite{Ryan}.   
IRA codes and e-IRA codes have the low decoding complexity characteristic of LDPC codes, and
the low encoding complexity characteristic of turbo codes \cite{McEliece, Ryan, Berrou}.  

LDPC error floors are caused by connected sets of cycles called
``stopping sets'' \cite{Di}. Codes with larger stopping sets generally have lower error
floors.   The design technique in \cite{Tian} attempts to maximize stopping set size by
maximizing the average number of connections leading outside small cycles, referred to as the 
ACE distance  $d_{ACE}$; simulations showed that LDPC codes with larger $d_{ACE}$
had lower error floors.  More recently, the authors of \cite{Lee} proposed a method of
directly estimating the variable and check nodes in the smallest stopping sets, along
with a code design algorithm to directly maximize the size of these sets.  The design
algorithm in \cite{Lee} resulted in codes with  significantly lower error floors 
than those designed according to \cite{Tian}.

The contribution of the present paper is a method of designing serially concatenated
IRA codes that achieve lower error floors than single IRA codes of equivalent 
rate and block size.  Two systematic component codes, with block length and rate equal to
the square roots of those of a comparable full-length IRA code, are connected in series, 
with an interleaver
between them.  This architecture is similar to that of turbo product
codes \cite{Pyndiah}, except that, rather than employing the
row-column interleaver of product codes, we design the interleaver
to avoid the convergence problems that lead to error floors.
We use the method of \cite{Lee} to estimate the stopping
sets of the component codes.  Then the stopping set data is used to design
the interleaver so that, as much as possible, stopping
set error events of one of the component codes are not mapped into stopping set
variable nodes of the other code.  Since each component code has the ability to successfully
decode the other code's non-convergent blocks, convergence problems are greatly reduced,
resulting in a lowered error floor at high SNR.  Because of the IRA
component codes, the concatenated system has relatively low encoding complexity compared
to a general irregular LDPC code.
The decoding complexity is about twice that of the comparable full-length IRA code,
due to the need for outer iterations between the component codes.

This paper is organized as follows.  Section \ref{sec: codec} summarizes the encoder and
decoder architectures.  Section \ref{sec: interleaver} presents the interleaver design.
Section \ref{simres} presents simulation results, and section \ref{sec: conc} concludes
the paper.

\section{Concatenated IRA Encoder and Decoder}
\label{sec: codec}

\begin{figure}[tbh]
\centerline{\epsfig{figure=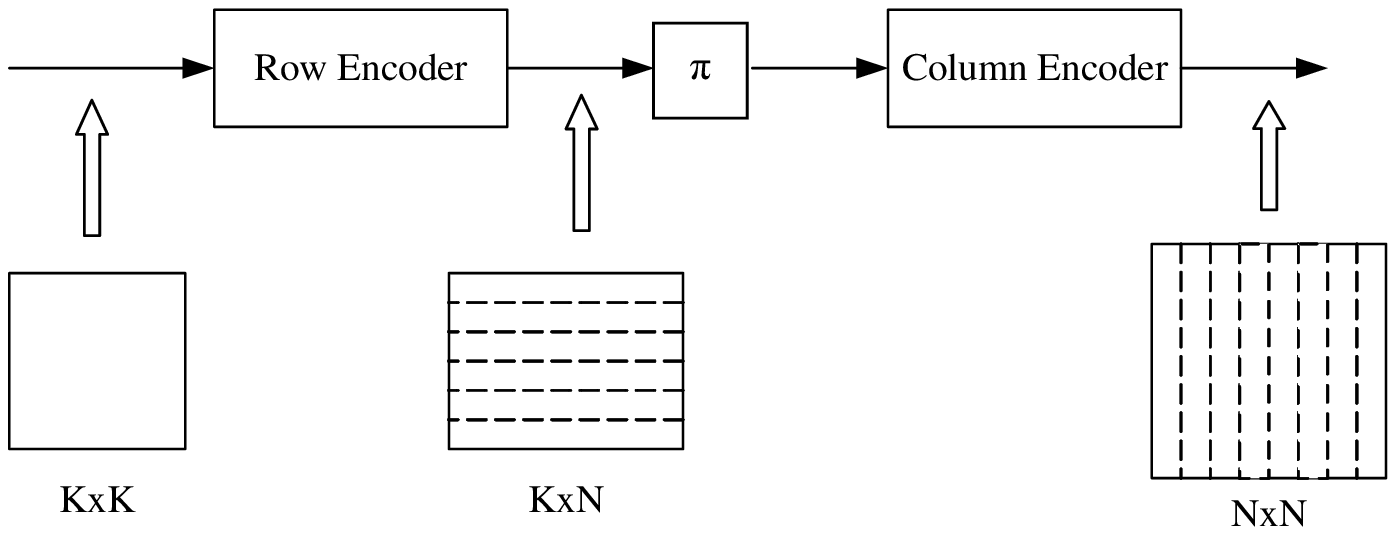, width=3.2in}}
\vspace{0.1in}
 \caption{Block diagram of the concatenated encoder with systematic IRA component codes connected by  interleaver (denoted by $\pi$).}
 \label{fig: eblock}
\end{figure}

A block diagram of the concatenated encoder is shown in Fig.~\ref{fig: eblock}. It consists of two systematic IRA component codes connected by an interleaver (denoted by $\pi$).
In the following discussion, we visualize the concatenated system as a
product code, with the two encoders operating on rows and columns.
The source data is arranged in a two-dimensional block of size $K \times K$.
The rows of the source block are first encoded with the outer $[N,K]$ systematic IRA code, 
yielding a $K \times N$ coded block in which the first $K$ elements of each row are 
systematic bits. Then the $K \times N$ coded block is passed through
the interleaver. The purpose of the interleaver is to minimize the intersection between
the stopping set error events of the row and column component codes. 
After the interleaver, each $K$-bit column is encoded with the inner 
$[N,K]$ systematic IRA code, producing an $N \times N$ codeword block.
The overall code rate is $R=K^2/N^2$.
The identical variable-node degree distributions of the two component codes are chosen to 
optimize their performances in the waterfall region according to the design algorithm given
in \cite{McEliece}, subject to the constraint that all weight-2 columns appear
in the $H_2$ section of the parity check matrix; the constraint helps lower the
error floors of the component codes.  All example codes designed in this paper used a fixed
check node degree of 10.  The variable-to-check node connections in
the component codes are optimized using the ACE algorithm of \cite{Tian}, in order to
further lower the error floors.
In our examples, the variable-to-check node connections in the
component codes are different, so that the codes have different stopping sets; 
however, the interleaver design
described in section \ref{sec: interleaver} also works if the component codes are
identical.  

\begin{figure}[tbh]
\centerline{\epsfig{figure=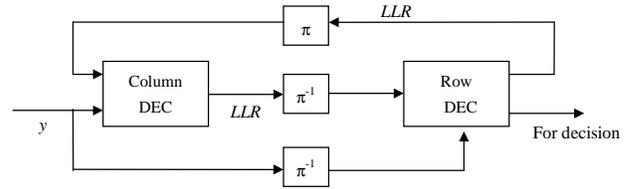, width=3.2in}}
\vspace{-0.1in}
 \caption{Block diagram of the concatenated decoder.}
 \label{fig: dblock}
\end{figure}

The decoder for the concatenated system employs iterative message passing 
between the decoders for the two component codes. 
The decoder block diagram is shown in Fig.~\ref{fig: dblock}. It consists of column
and row decoders connected by the interleaver and de-interleaver. 
The received channel data is decoded column by column by a standard $[N,K]$ IRA decoder
employing the sum-product algorithm (SPA, \cite{spa-paper}) on the code's Tanner
graph; the column decoder uses the extrinsic
information from the row decoder as {\em a priori} information.  The column decoder outputs
a $K \times N$ block of extrinsic information LLRs. 
The column decoder's output extrinsic information is then passed through the
interleaver and used as prior information by the row decoder. 
The row decoder makes use of the de-interleaved channel information and the 
prior information to decode the data row by row, and outputs a $K \times N$ block
of LLRs to be used for final decoding decisions, 
along with a $K \times N$ block of extrinsic
LLRs for the column decoder to use during the next iteration.  

\section{Interleaver Design}
\label{sec: interleaver}
The reasons to encode/decode using the structure described above rather
than using a single $[N^2,K^2]$ IRA code are as follows.   
The performance of an LDPC code at high SNR (i.e., in the error floor region)
is not determined by the code's minimum distance, but rather by sets of interconnected
short cycles (called stopping sets) that prevent the decoder from converging to a 
valid codeword.  If we can design the interleaver to prevent the
mapping of stopping set error events from one of the component codes into stopping
set nodes of the other code, then the concatenated structure will 
help improve the performance at high SNR.     

The definition of a stopping set used in this paper is as follows.  
A variable-node set is called a stopping set if all its neighbors are connected to this 
set at least twice \cite{Tian}. In LDPC codes at high SNR, error events occur on the 
smallest stopping sets with higher probability than on larger stopping sets or 
non-stopping sets. 
To simplify, if a variable node is a part of a stopping set, we call it a sensitive node.  

Here is an example of how an error event from one IRA component
code could propagate into the 
other one.  Suppose variable nodes $(6,9,25)$ are sensitive nodes of the column component 
code and that errors occur on these positions.
Since each column uses the same component code, errors will occur on these positions 
on most columns, i.e., at the end of column decoding, most positions of rows $(6,9,25)$ 
are errors. If we do nothing but directly input these rows to the row decoder, 
the outputs will have a large number of errors (perhaps even larger then the number of
input errors) due of the bad prior information.  If we pass the output extrinsic 
information from the column decoder through an interleaver before it is fed to the 
row decoder, the errors will not be concentrated on rows $(6,9,25)$ and hence can
be corrected more easily.
Therefore, we postulate two interleaver design rules for the concatenated system:
\begin{enumerate}
\item Spread concentrated errors all over the data block.
\item Avoid mapping the sensitive nodes of the row (column) component code into the 
sensitive nodes of the column (row) component code.
\end{enumerate}

\begin{figure}[tbh]
\centerline{\epsfig{figure=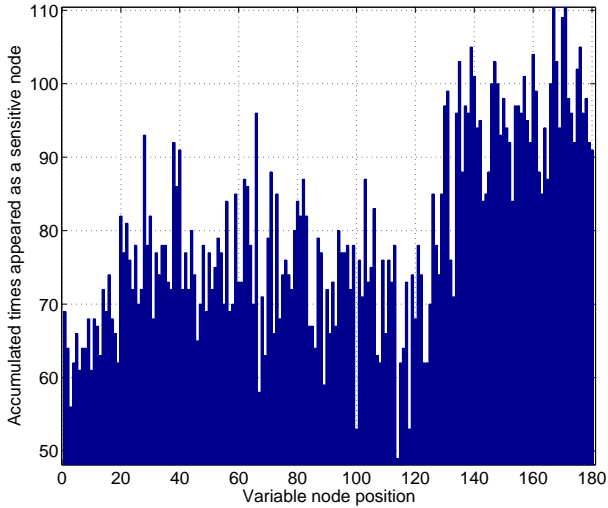, width=3.6in}}
\vspace{-0.1in}
 \caption{Sensitivity measurement via stopping set detection. 
The sensitivity counts on the vertical axis
are accumulated by running the algorithm of \cite{Lee} on every possible starting
variable node, and then counting the number of times any given node appears in
the detected stopping sets.} 
 \label{fig: STPexample}
\end{figure}

The sensitive positions of a component code can be determined by
employing the stopping set detection algorithm of \cite{Lee}. 
For a given starting variable node, the algorithm in \cite{Lee} finds a stopping set
containing that node, but does not guarantee that the detected set is minimal; 
thus, some relatively less-sensitive nodes may be included in the set. 
To find the most sensitive nodes, we repeatedly run the detection algorithm by starting
from every variable node in the code, and count
the accumulated times each node appears in a stopping set; the higher the
count, the more sensitive the node. 
Fig.~\ref{fig: STPexample} shows the results of running the algorithm of \cite{Lee} 
over the $[181,128]$ row component IRA code by starting from each variable node.  
In Fig.~\ref{fig: STPexample}, the maximum sensitivity count is $181$.   It is clear
from the figure that some nodes are highly sensitive, and that most of the parity bits
(bits 129-181) have high sensitivity counts.

Based on the above design rules, we design the interleaver 
by starting with a random interleaver and imposing additional constraints. 
First, a relatively good $K \times N$ random interleaver is found by simulation. 
Then the stopping sets of the row and column component codes are detected 
using the method of \cite{Lee}.
For given sensitive nodes $i$ and $j$ of the row/column component codes 
$i \in \{I_0,I_1,\cdots,I_n\}$ and $j \in \{J_0,J_1,\cdots,J_m\}$, where 
$\{I_0,I_1,\cdots,I_n\}$ and $\{J_0,J_1,\cdots,J_m\}$ are the sensitive nodes
of the row and column component codes respectively, 
we modify the random interleaver so that no element in the $j$th row 
before passing through the interleaver is located in the $i$th column 
after passing through the interleaver.  If the random interleaver maps any
element in row $j$ to column $i$ (the ``bad mapping'' condition), 
then that element is re-mapped to a random position in the output block, and
the element formerly at that random position is mapped into the position of
the element in row $j$; this re-mapping continues until either no bad mappings
are found or all the possible positions in the interleaver have been checked,
in which case no interleaver solution is possible. 
Since the stopping set detection algorithm yields a large set, 
we select only the most sensitive nodes (i.e., the nodes with highest sensitivity
counts in a histogram like that of Fig.~\ref{fig: STPexample}) 
to design the interleaver at the beginning.  Then we increase the 
number of selected sensitive nodes step by step until we cannot 
find a solution for the interleaver.  

\section{Simulation Results}
\label{simres}

\begin{figure}
\centerline{\epsfig{figure=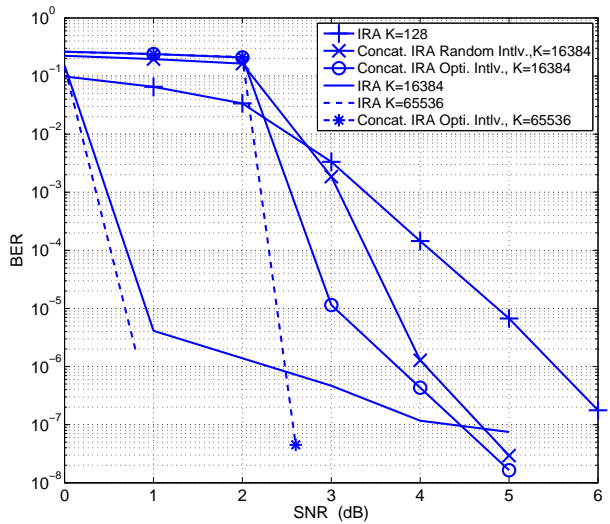, width=3.6in}}
\vspace{-0.1in}
 \caption{Simulation results.  All codes are rate 1/2, except for
the $K=128$ IRA code, which is rate 0.707.}
 \label{fig: SimuResults}
\end{figure}

The Monte Carlo simulation results for the proposed concatenated IRA code structure 
on the binary-input AWGN channel are 
shown in Fig.~\ref{fig: SimuResults}.
In the figure, the right-most curve (marked by `+' symbols) is the
performance of a 
single IRA component code with source block length 
$K=128$ bits and code rate $0.707$. 
The second rightmost curve (marked `x') is the proposed concatenated 
code with block size $K^2=16384$, rate $0.5$, and a random 
interleaver;
the random interleaver was found by (non-exhaustive) search 
over a large
number of randomly generated interleavers.
The solid line with circle markers is the same code structure as the second curve, 
but uses an interleaver based on the design rules proposed in section \ref{sec: interleaver};
this designed interleaver used the random interleaver of the `x' curve as the
design's starting point.
The dashed line with star markers is the BER of a rate 1/2, block length $K^2=65536$ 
concatenated IRA code with optimized interleaver.  
For comparison, we also simulated single long block length IRA codes with rate $0.5$;
the solid line is with source block length $16384$ and the dashed line is with source 
block length $65536$.

All the single IRA code simulations were run until either a valid codeword was decoded,
or 100 iterations were performed. 
For both the 16384-bit concatenated curves the decoder was run for a
total of 10 outer iterations between the component codes, and the component
codes were each iterated 10 times per outer iteration.  Component decoding
(on a given row or column) was terminated before 10 iterations if a valid codeword was decoded.
The concatenated iteration schedule
was determined experimentally, and therefore may not be optimal.  (Further optimization
of the iteration schedule using, e.g., EXIT charts \cite{tenBrink}, will be
the focus of future work.)   The complexity
of the 16384-bit concatenated decoder is thus approximately twice that of the 16384-bit
single IRA code, although at higher SNR the complexity of the concatenated system is
relatively higher because termination events for the concatenated code eliminate only
single rows or columns from the iteration, not the entire codeword.  
The 65536-bit concatenated decoder was run for a total of 10 outer iterations 
with 20 inner iterations
per outer iteration, so its decoding complexity is about four times that of the
single 65536-bit IRA code.  

From the figure it is clear that, 
although the concatenated 16384-bit IRA code has an SNR penalty in 
the waterfall region (about 2.1 dB SNR at BER $10^{-5}$) compared to the 
single 16384-bit IRA code of equivalent rate,
it has a much lower error floor.  There is a crossover point between these two 
codes' BER curves at a BER of about $10^{-7}$, and the BER of the concatenated IRA code 
decreases much faster than that of the single IRA code at high SNR. 
By comparing the 16384-bit concatenated codes' performance with different interleavers, 
we see that the proposed interleaver design can achieve
significant gains 
(about $0.7$ dB at $10^{-5}$ and $0.3$ dB at $10^{-7}$) over the random interleaver
used as the design starting point, 
which means the idea of separating the component codes' stopping sets works.

The $K=128$ example component codes are quite short.
We conjecture that when the block length is increased the penalty in the waterfall
region will decrease, since the component IRA codes will asymptotically approach
capacity as the block length increases.  
This conjecture is partly supported by the smaller SNR penalty (about 1.7 dB at BER $10^{-5}$)
of the 65536-bit rate-1/2 concatenated code compared to the equivalent-rate 65536-bit 
IRA code, although part of the improvement over the 16384-bit codes 
may be due to the increased decoder iterations
allocated to the 65536-bit concatenated system.

\section{Conclusions}
\label{sec: conc}
This paper has demonstrated that serial concatenation of two IRA codes connected
by an appropriately designed interleaver can greatly lower the level and slope
of the BER curve in the high SNR region, compared to a single IRA code of
equivalent length and rate.  
We believe that the proposed approach will also work 
\vspace*{2.5in}
\pagebreak
with more general LDPCs as component codes, including, e.g., e-IRA codes or 
codes optimized with the error-floor lowering algorithm of \cite{Lee}.
Future work will focus on reducing the SNR penalty of the concatenated codes
in the waterfall region through more rigorous 
optimization of the iteration schedule, and through use of longer block
length component codes.

\section*{Acknowledgment}
The authors would like to thank the National
Science Foundation for providing support
for the work presented in this paper, under 
grant CCF-0635390.



%

\end{document}